\documentclass[english,aps,preprint]{revtex4}
\usepackage[T1]{fontenc}
\usepackage[latin9]{inputenc}
\usepackage{textcomp}
\usepackage{graphicx}

\makeatletter

\providecommand{\tabularnewline}{\\}

\@ifundefined{textcolor}{}
{%
 \definecolor{BLACK}{gray}{0}
 \definecolor{WHITE}{gray}{1}
 \definecolor{RED}{rgb}{1,0,0}
 \definecolor{GREEN}{rgb}{0,1,0}
 \definecolor{BLUE}{rgb}{0,0,1}
 \definecolor{CYAN}{cmyk}{1,0,0,0}
 \definecolor{MAGENTA}{cmyk}{0,1,0,0}
 \definecolor{YELLOW}{cmyk}{0,0,1,0}
}

\makeatother

\makeatother

\usepackage{babel}

\usepackage{cancel}

\makeatother

\usepackage{babel}
\begin{document}

\title{Resonant Production of Sbottom via RPV Couplings at the LHeC}

\author{S. Kuday}

\email{kuday@science.ankara.edu.tr, kuday@cern.ch}

\selectlanguage{english}%

\affiliation{Institute of Accelerator Technologies, Ankara University, Ankara,
Turkey}

\email{http://hte.ankara.edu.tr}

\selectlanguage{english}%
\begin{abstract}
Resonant production of scalar bottom, which is allowed in R-parity
violating interactions of supersymmetry, has been investigated at
the LHeC collider. Although searching for the physics beyond the standard
model is a primary task of the LHC program, recently, an $\mathit{e^{-}p}$
collider (LHeC) is proposed to complement and resolve the observation
of new phenomena at the TeV scale. In this paper, we have studied
on the prospects of improving constraints for $\hat{L}\hat{Q}\hat{D}$
couplings $\lambda_{ijk}'$ through the process $\mathit{e^{-}+p\rightarrow\widetilde{b^{*}}\rightarrow\mu^{-}+\overline{q}}$
where $\mathit{q}$ denotes the up type quarks. It is shown that constraints
on $\lambda_{ijk}'$ can be improved up to $10^{-3}$ for $1\: fb^{-1}$
integrated luminosity at 95\% C.L. with 60 GeV $\mathit{e^{-}}$ beam
option of the LHeC. 
\end{abstract}

\pacs{14.80.Ly, 11.30.Pb, 12.60.Jv \vspace{0.5cm}}

\maketitle

\section{Introduction}

Theoretical structure of supersymmetry (SUSY), which is recently an
active area of research and interest at the LHC, allows gauge-invariant
and renormalizable interactions that violate the conservation of lepton
and baryon numbers. In the framework of minimal supersymmetric standard
model (MSSM), these interactions are forbidden by imposing an additional
global symmetry that leads to the conservation of a multiplicative
quantum number: R-parity \cite{Barbier}, which is defined as $R=(-1)^{3(B-L)+2S}$,
where $B,\, L$ and $S$ are the baryon number, lepton number and
spin, respectively. As a natural consequence of this phenomenology,
all the SM particles and Higgs boson have even R-parity $(R=+1)$,
while all the sfermions, gauginos and higgsinos have odd R-parity
$(R=-1)$. One of the highest motivations for R-parity conserved MSSM
is that it provides sparticles to be produced in pairs since two odd
particles always give even number of R-parity. Although no SUSY signal
has been detected yet, pair production of sparticles may be an important
clue for final states in SUSY searches at the LHC. 

From the theoretical grounds of SUSY, one could expose that R-parity
conservation is actually inherited from the conservation of $B$ and
$L$ quantum numbers which is the natural consequences of a renormalizable
and a Lorentz invariant theory. For such an extended SM theory, it
is not necessary to keep those variables still conserved as long as
the algebraic structure is safe. Although non-conservation of both
$B$ and $L$ quantum numbers leads to rapid proton decay, a firm
restriction to RPV (R-Parity Violating) couplings guarantees a stable
proton. Furthermore, allowing many of the interactions with the sparticles
in the RPV SUSY model provides even richer phenomenology comparing
to the other models. However, many of the interactions in these terms
may appear to be strictly supressed in the nature. Thus, practical
application of RPV in MSSM also reveals several implications: firstly,
sparticles can be produced in resonance processes as well as in pairs
and secondly stabilization of particles (e.g.: dark matter) may not
be guaranteed directly. It has been showed that in the context of
bilinear RPV model both gravitino \cite{GRAVITINO1,GRAVITINO2} or
axino \cite{AXINO} as dark matter are consistent with a lifetime
exceeding the age of universe. Considering the neutrino issue in the
SM, bilinear RPV which is induced by bilinear terms in the superpotential
can explain the neutrino masses and mixings in compatible with the
current data without invoking any GUT-scale physics \cite{NEUTRINO}. 

From the recent experimental data, the highest constraints for gluino
mass reached about 1.5 TeV with 95\% C.L. in GMSB and CMSSM searches
at $\sqrt{s}=8\: TeV$ according to the ATLAS \cite{ATLAS} and CMS
\cite{CMS} results. For stop and sbottom masses, recent constraints
are $m(\tilde{t})>660\: GeV$ \cite{ATLAS2} for $L=20.5\: fb^{-1}$
and $ $$ $$m(\tilde{b})>620\: GeV$ \cite{ATLAS3} for $L=12.8\: fb^{-1}$
at the LHC. Ongoing researches that can be interpreted in the context
of R-parity violating supersymmetric scenarios at the LHC set the
limit $\tilde{q}>700\: GeV$ \cite{ATLAS4,CMS2} for squark masses
in muon + jets final states. 

As the continuation of the LHC physics program, LHeC \cite{LHeC1,LHeC2}
can extend these researches into the unexplored high mass regions
with a linac-ring configuration which has been decided in the CDR
\cite{LHeC3} to continue technical design work. It should be emphasized
that the parameter space which will be covered at the LHeC, also intersects
with LHC searches, so that the main motivation of this work will be
to compare limits between the LHeC as the future collider and LHC
as the recent collider and to search for a possibility to improve
those limits. After the LHeC starts running in full power, the first
task will be to reconsider those limits and improve them to constraint
R-parity violating squarks. Throughout this work, we will consider
the basic energy option as the main reference for LHeC, namely, $e^{\pm}$
= 60 GeV and $p$ = 7 TeV with $L\,=\,10^{33}\: cm^{-2}s^{-1}$.

\section{S\i{}gnal Production and Decay via RPV Interactions}

\begin{figure}
\includegraphics[scale=0.9]{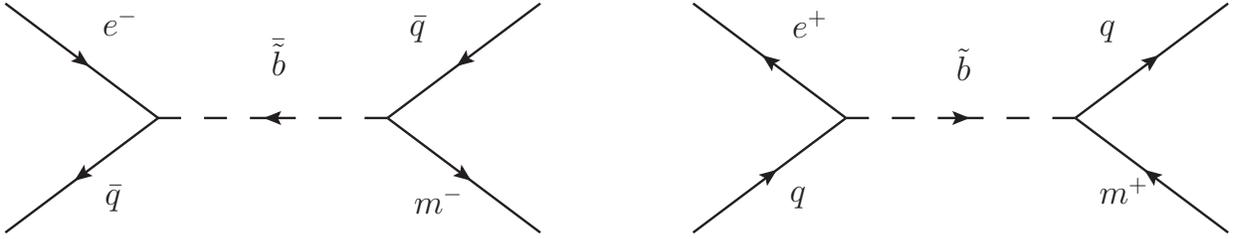}

\caption{Feynman diagrams of signal production where $q=u,c$ quarks\label{fig:1}}
\end{figure}
The R-parity violating extension of the MSSM superpotential is given
by

\begin{equation}
W_{RPV}=\frac{1}{2}\lambda_{ijk}\epsilon^{ab}L_{i}^{a}L_{j}^{b}\overline{E}_{k}+\lambda'_{ijk}\epsilon^{ab}L_{i}^{a}Q_{j}^{b}\overline{D}_{k}+\frac{1}{2}\lambda''_{ijk}\epsilon^{\alpha\beta\gamma}\overline{U}_{i}^{\alpha}\overline{D}_{j}^{\beta}\overline{D}_{k}^{\gamma}\label{eq:1}
\end{equation}
where $i,j,k=1,2,3,4$ are the family indices; $a,b=1,2$ are the
$SU(2)_{L}$ indices and $\alpha,\beta,\gamma$ are the $SU(3)_{C}$
indices. $L_{i}(Q_{i})$ are lepton (quark) $SU(2)$ doublet superfields;
$E_{i}(D_{i},U_{i})$ are the charged lepton (down-type and up-type
quark) $SU(2)$ singlet superfields. The couplings $\lambda_{ijk}$
and $\lambda''_{ijk}$ correspond to the lepton number violating and
baryon number violating couplings, respectively. One can easily see
that the $\lambda'_{ijk}$ coupling constants are antisymmetric under
the exchange of the first two indices and extract the $\lambda'_{ijk}$
part of the Lagrangian as; 
\begin{equation}
L_{\lambda'}=-\lambda'_{ijk}[d_{Rk}^{\dagger}\overline{\nu_{i}^{c}}P_{L}d_{j}+\tilde{d_{Lj}}\overline{d}_{k}P_{L}\nu_{i}+\tilde{\nu}_{i}d_{k}P_{L}d_{j}-\tilde{d_{Rk}^{\dagger}}\bar{e_{i}^{c}}P_{L}u_{j}-\tilde{e_{Li}}\bar{d}_{k}P_{L}u_{j}-\tilde{u_{Lj}}\bar{d_{k}}P_{L}e_{i}]+h.c.\label{eq:2}
\end{equation}
Here, the fourth term directly corresponds to the vertex factors of
the diagrams in Fig.\ref{fig:1}. So one can write the parton-level
differential cross section for signal in the rest frame of final muon
and quark states as;

\begin{equation}
\frac{d\sigma}{d\Omega}=\frac{(\lambda'_{123}\lambda'_{232})^{2}}{(16\pi)^{2}}\frac{\hat{s}}{(\hat{\hat{s}-m_{\tilde{b}}^{2})^{2}-(\Gamma m_{\tilde{b}})^{2}}}\label{eq:3}
\end{equation}
where $m_{\tilde{b}}$ is sbottom mass and $\Gamma$ is the total
width of sbottom that can be calculated as $(\lambda'_{ijk})^{2}m_{\tilde{b}}/8\pi$.
\begin{figure}
\includegraphics{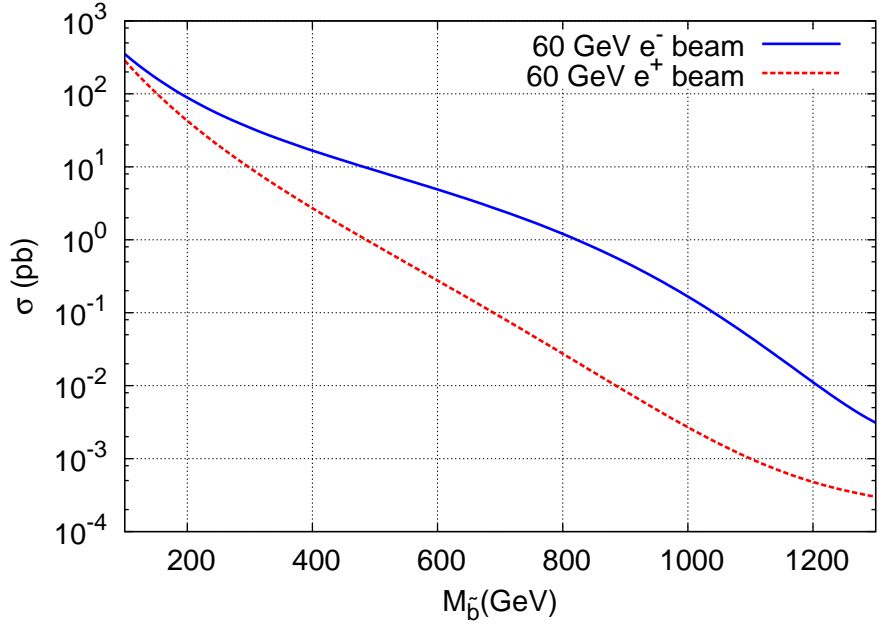}

\caption{Cross sections vs. sbottom mass\label{fig:2}}
\end{figure}

Since we take the single dominance hypothesis for granted, lighter
sbottom mass eigenstate will be the actual object here whenever we
refer to sbottom. For the signal production, one could immediately
calculate that the contributions of other down type scalar superpartners
are negligible and parton-level contributions of all other quarks
are minor except $u,c$ quarks. Therefore, we have taken into account
these contributions to evaluate the total cross sections as in Fig.\ref{fig:2}
using the COMPHEP \cite{COMPHEP} event generator and CTEQ6M PDF \cite{PDF}
package. In SUSY phenomenology, the magnitudes of the RPV couplings
are arbitrary, and they are restricted only from the phenomenological
considerations. Therefore two standard bounds are taken as \cite{BOUNDS};

\begin{equation}
\lambda'_{113}=\lambda'_{123}\leq0.18\,,\qquad\lambda'_{231}=\lambda'_{232}\leq0.45\label{eq:4}
\end{equation}

Here, it is worthwhile to emphasize that for electron and positron
beam options, calculations explicitly show that the $e^{-}$ beam
options always deliver the highest cross section values, even for
60 GeV $e^{-}$ beam option in the low mass region. This result seems
to be contrary with the stop resonance production at the LHeC \cite{StopLHeC}
where $e^{+}$ beam option delivers the higher cross section values.
The main reason of that difference is related with the subprocess
$\mathit{e^{-}+q\rightarrow\widetilde{b^{*}}\rightarrow\mu^{-}+\overline{q}}$
where $q$ denotes $u,c$ quarks whereas for stop production main
contribution comes from $b$ quarks in the initial state. Therefore,
equation {[}\ref{eq:3}{]} yields to stronger signal values than that
of the stop resonance production. For the rest of this work, we choose
60 GeV $e^{-}$ beam option as the default option for investigating
kinematical distributions and exclusion limits.

\begin{figure}
\includegraphics{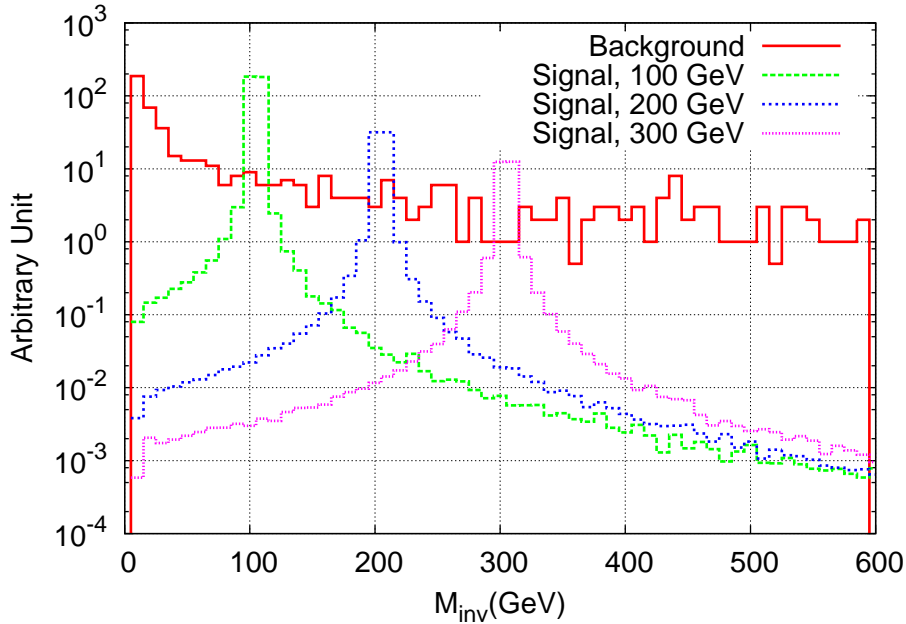}

\caption{Invariant mass distribution of signal and background \label{fig:3}}
\end{figure}

\begin{figure}
\includegraphics{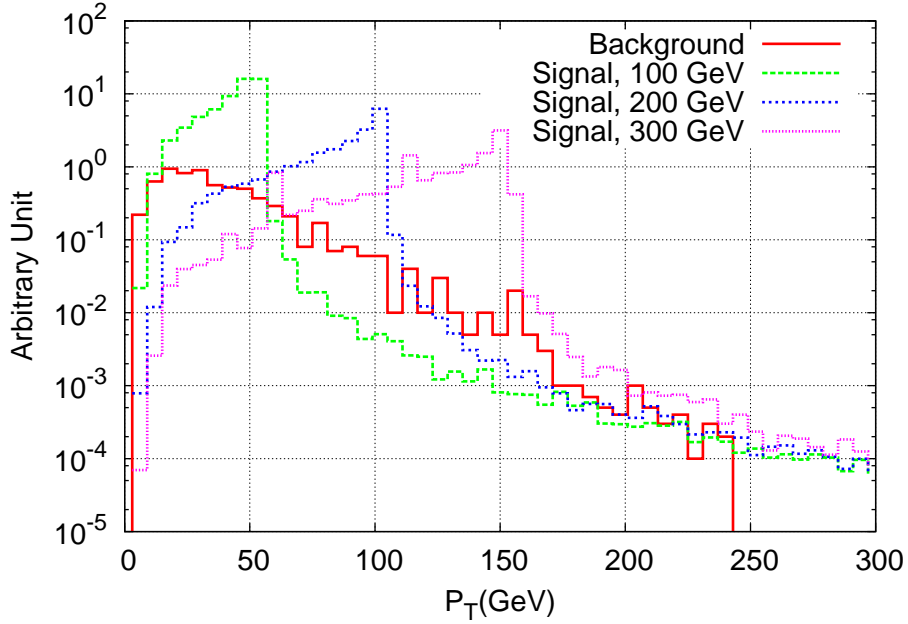}

\caption{Transverse momentum for jets\label{fig:4}}
\end{figure}

\begin{figure}
\includegraphics{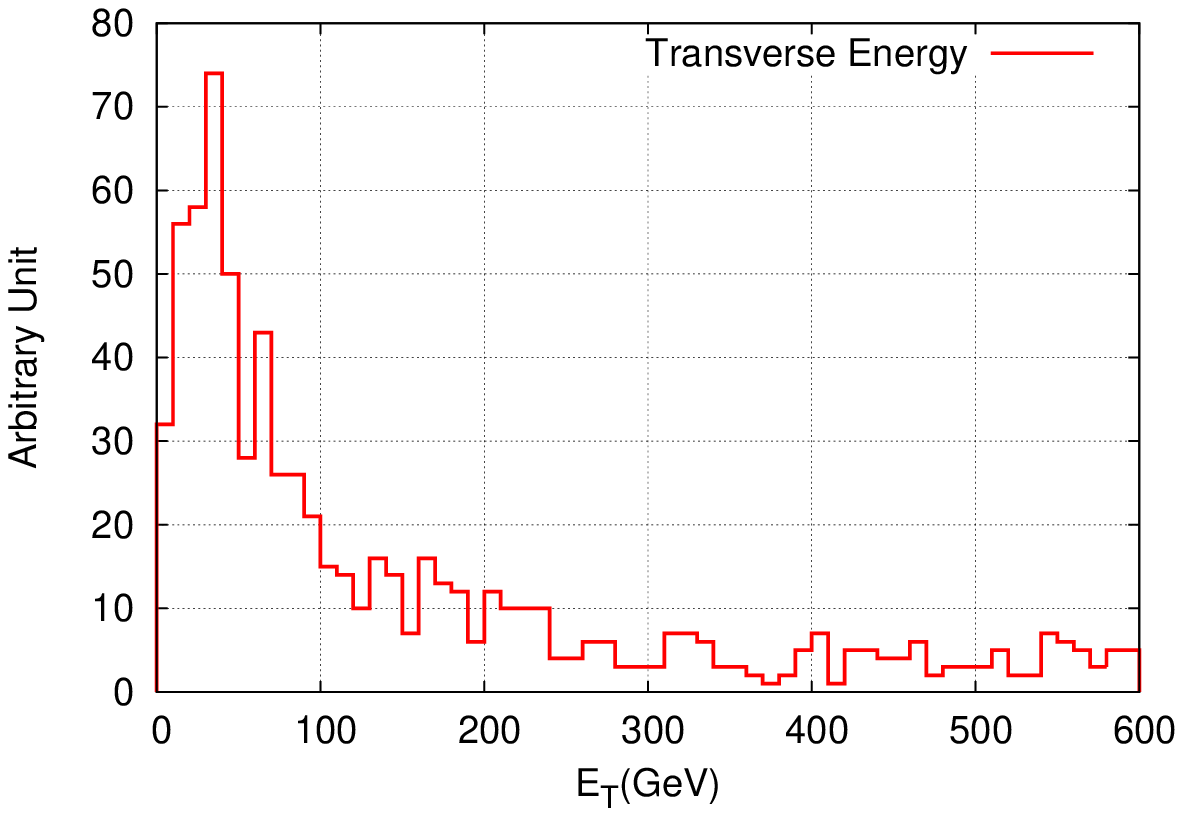}

\caption{Missing transverse energy of the total background \label{fig:5}}
\end{figure}

\section{Background Processes}

The process $e^{\pm}+p\rightarrow\mu^{\pm}+q/\bar{q}+X$ where $q$
denotes $u,c$ quarks seems to be the main background resource for
both beam options at the LHeC. The reducible SM background comes through
the subprocess $e^{-}+p\rightarrow\nu_{e}+q/\bar{q}+W^{-}$ where
W boson rapidly decays via $\mu^{-}\bar{\nu}_{\mu}$ channel. Note
that, vetoing $b/\bar{b}$ quark contributions in the final state
reduces these subprocesses for considerable amount. In experimental
point of view, the background may even be reduced more if c-tagging
option implemented for the final state quarks. In our case, we didn't
take into account any c-tagging options because of the considerably
low tagging efficiencies. We have obtained the comparisons of kinematic
distributions for backgrounds and for RPV signals using PYTHIA 6.4
\cite{PYTHIA 6.4} and COMPHEP \cite{COMPHEP} respectively as in
Fig.\ref{fig:3} and Fig.\ref{fig:4}. We have built a new model implementing
RPV interactions and vertex factors in COMPHEP for signal while we
have used SM event generator of PYTHIA 6.4 \cite{PYTHIA 6.4} for
background normalizing over $10^{4}$ events. $P_{T}$ distributions
of the jets in the final states will be naturally at the order of
half sbottom mass since outgoing particles, muon and jets are back-to-back
in the transverse plane neglecting the missing transverse energy.
Since we have neutrinos that will escape from detection in the final
states leaving a significant missing transverse energy, it is important
to have non-zero $\cancel{\it{E}}_{T}$ distribution as in Fig.\ref{fig:5}.

\section{Event Selection and D\i{}scussion }

\begin{table}
\begin{tabular}{|c|c|c|c|c|}
\hline 
$M_{\tilde{b}}$ (GeV) & $\sigma(e^{-}p)$ $(pb)$ & Required $L_{int}(e^{-}p)$$(pb^{-1})$ & $\sigma(e^{+}p)$ $(pb)$ & Required $L_{int}(e^{+}p)$$(pb^{-1})$\tabularnewline
\hline 
\hline 
100 & 211.78  & 0.022  & 201.38  & 0.023 \tabularnewline
\hline 
200 & 38.39  & 0.154  & 22.17  & 0.318 \tabularnewline
\hline 
300 & 13.71  & 0.492  & 4.85  & 2.5 \tabularnewline
\hline 
400 & 6.82  & 1.192  & 1.35  & 18.836 \tabularnewline
\hline 
500 & 3.76  & 2.683  & 0.41  & 157.761 \tabularnewline
\hline 
600 & 2.1 & 5.707  & 0.13  & $1.382\times10^{3}$\tabularnewline
\hline 
700 & 1.11  & 13.547  & $3.78\times10^{-2}$ & $1.321\times10^{4}$\tabularnewline
\hline 
800 & 0.53 & 32.010  & $1.3\times10^{-2}$ & $7.3\times10^{4}$\tabularnewline
\hline 
900 & 0.21 & 79.491  & $2.1\times10^{-3}$  & $11.738\times10^{5}$\tabularnewline
\hline 
1000 &  $6.55\times10^{-2}$ & 217.423 & $3\times10^{-4}$ & $14.37\times10^{6}$\tabularnewline
\hline 
\end{tabular}\caption{Required luminosities and cross sections of sbottom at the 60 GeV
$e^{\pm}$ beam option of LHeC for 95\% C.L. with $\lambda'_{113}=\lambda'_{123}=0.18$
and $\lambda'_{231}=\lambda'_{232}=0.45$\label{tab:1} }
\end{table}

\begin{figure}
\includegraphics{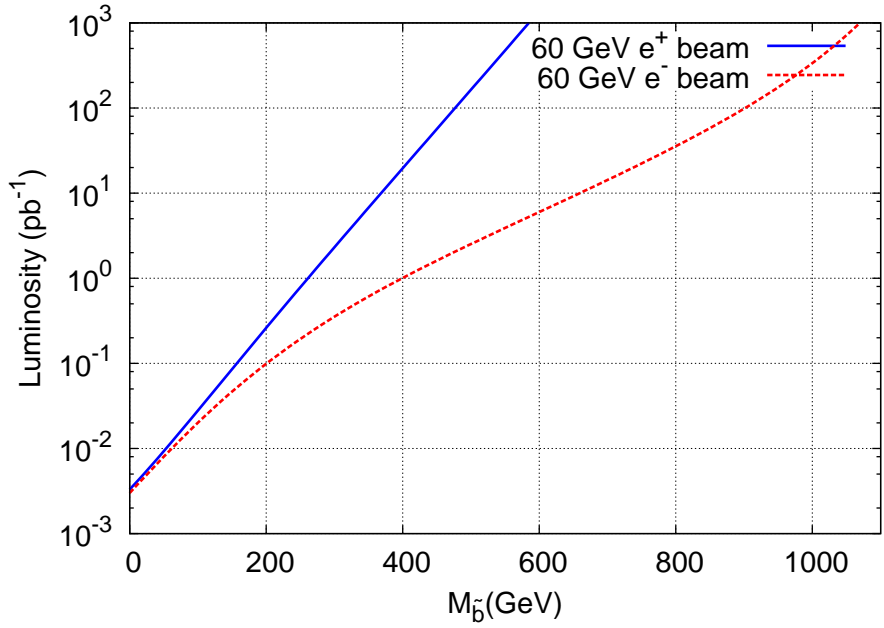}

\caption{Integrated luminosity vs. sbottom mass for 95\% C. L.\label{fig:6} }
\end{figure}

\begin{figure}
\includegraphics{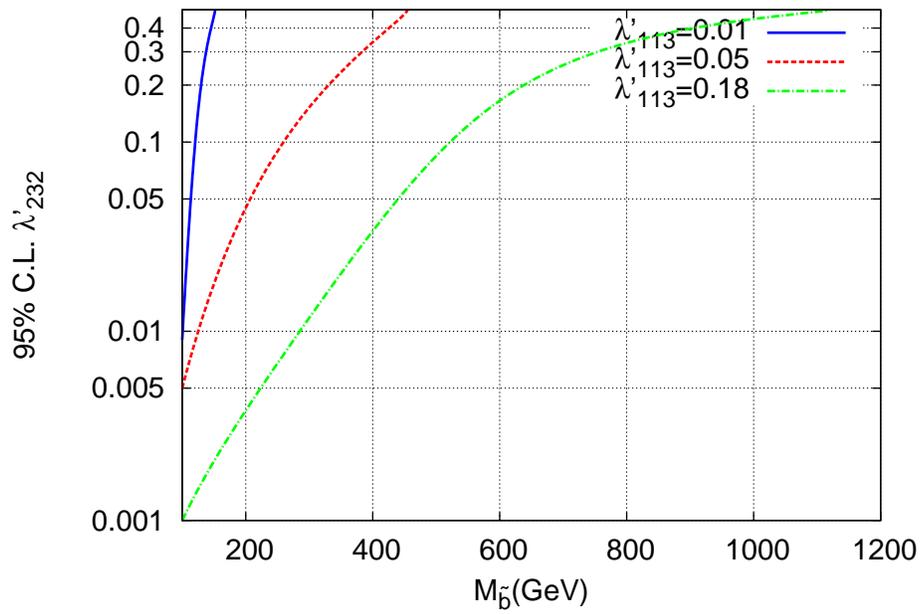}

\caption{Attainable limits for the sbottom mass and RPV couplings at 60 GeV
$e^{-}$ beam option of LHeC \label{fig:7}}
\end{figure}

For event selection part of the analysis, a strict strategy has been
introduced before for RPV resonance particles \cite{StopLHeC} in
order to reduce large SM background. In our case, we have developed
the following cuts and optimizations:
\begin{itemize}
\item Kinematic cuts: for muons $p_{T}^{\mu}>25$ GeV and $ $$\left|\eta_{\mu}\right|<2.5$;
for jets $p_{T}^{q}>25$ GeV and $ $$\left|\eta_{q}\right|<3.5$.
\item Missing transverse energy veto: $\cancel{\it{E}}_{T}$ < 25 GeV.
\item Invariant mass cut: $M_{\mu q}>85$ GeV and mass window cut in accordance
with the energy resolution.
\item Vetoing b-jets with 60\% efficiency: Same assumption for b-jet identifi{}cation
in experiments since we need to identify b-jets before vetoing. 
\end{itemize}
After implementing above selection criteria, background cross sections
are calculated as 1.7 fb for 60 GeV $e^{-}$ beam option and 1.6 fb
for 60 GeV $e^{+}$ beam option. For signal production, events always
survived not below than 85\% for sbottom masses between 100 and 1000
GeV. In Table.\ref{tab:1}, one can see the required luminosities
to reach $2\sigma$ significance value (95\% C.L.) for both 60 GeV
$e^{-}$ beam and 60 GeV $e^{+}$ beam options. In significance calculations,
we have always used $S/\sqrt{S+B}$ formulation where $S$ number
of signal and $B$ number of background events. It is obvious that
LHeC can exclude sbottom mass up to 1000 GeV in its first runs with
217.5 $pb^{-1}$ integrated luminosity if there is no apparent excess
from SM predictions on $\mu$ + jets fi{}nal states. Likewise, we
depicted an extended plot of Table \ref{tab:1} in Fig.\ref{fig:6}.
Attainable limits for sbottom mass with respect to RPV couplings $\lambda'_{113}=\lambda'_{123}$
and $\lambda'_{231}=\lambda'_{232}$ are presented in Fig.\ref{fig:7}
at 60 GeV $e^{-}$ beam option of LHeC for $1\: fb^{-1}$ integrated
luminosity. One can see here that LHeC can exclude sbottom mass up
to 1200 GeV. The main reason of $\lambda'_{113}=0.18$ line extending
to high mass region is a few survived backgrounds events after selection
criteria. With respect to the recent limits of sbottom mass (\textasciitilde{}
700 GeV), minimum attainable limits to RPV couplings $\lambda'_{232}$
calculated at around $0.25$ for a fixed $\lambda'_{113}=0.18$ value.

\section{Conclusion}

In this study, we introduced a phenomenological approach for constraining
$\hat{L}\hat{Q}\hat{D}$ couplings via RPV $\mathit{e^{-}+p\rightarrow\widetilde{b^{*}}\rightarrow\mu^{-}+\overline{q}}$
process ($q=u,c$ ). Resonance production of sparticles via RPV processes
is a great advantage for obtaining a stronger signal although the
specific final states can broaden the total background just as in
our case for $\mu$ + jets fi{}nal state. We implemented a stricter
event selection in the limits of experimental capabilities to optimize
the sensitivity of signal. Considering 60 GeV $e^{\pm}$ beam options
of LHeC, we presented the cross sections and required luminosites
at 95\% C. L. for $\lambda'_{113}=\lambda'_{123}=0.18$ and $\lambda'_{231}=\lambda'_{232}=0.45$
and attainable limits for sbottom mass with respect to RPV couplings.
In conclusion, LHeC can extend the exclusion limits of $\hat{L}\hat{Q}\hat{D}$
couplings up to $10^{-3}$ for $1\: fb^{-1}$ integrated luminosity
at 95\% C.L. with 60 GeV $\mathit{e^{-}}$ beam option.
\begin{acknowledgments}
I would like to express my greatest gratitude to Saleh Sultansoy for
his fruitful discussions and helpful comments. This work is partially
supported by Turkish Accelerator Center (TAC) Project, Turkish Atomic
Energy Authority (TAEK) and T.R. Ministry of Development (DPT). \end{acknowledgments}

\end{document}